\documentclass[aps,twocolumn,amsfonts,amsmath,amssymb,prb]{revtex4}
\usepackage{graphicx}
\usepackage{graphics}

\usepackage{amsfonts}
\usepackage{amssymb}
\usepackage{bm}
\newcommand{\bmu}{\boldsymbol{\mu}}

\begin{document}

\title{From molten salts to room temperature ionic liquids: Simulation studies on chloroaluminate systems}

\author{Mathieu Salanne$^1$}
\author{Leonardo J. A. Siqueira$^2$}
\author{Ari P. Seitsonen$^3$}
\author{Paul A. Madden$^4$}
\author{Barbara Kirchner$^5$}
\affiliation{$^{1}$~UPMC Univ Paris 06, CNRS, ESPCI, UMR 7195, PECSA, F-75005 Paris, France. Fax: +33 1 4427 3228; Tel: +33 1 4427 3265; E-mail: mathieu.salanne@upmc.fr}
\affiliation{$^{2}$~Laborat\'orio de Materiais H\'ibridos, Universidade Federal de S\~ao Paulo, CEP 04024-002, S\~ao Paulo, SP, Brazil}
\affiliation{$^{3}$~Physikalisch-Chemisches Institut, University of Zurich, Winterthurerstrasse 190, CH-8057 Zurich, Switzerland}
\affiliation{$^{4}$~Department of Materials, University of Oxford, Parks Road, Oxford OX1 3PH, UK.}
\affiliation{$^{5}$~Wilhelm-Ostwald Institute of Physical and Theoretical Chemistry, University of Leipzig, Linn\'estr. 2, D-04103 Leipzig, Germany}

\begin{abstract}
An interaction potential including chloride anion polarization effects, constructed from first-principles calculations, is used to examine the structure and transport properties of a series of chloroaluminate melts. A particular emphasis was given to the study of the equimolar mixture of aluminium chloride with 1-ethyl-3-methylimidazolium chloride, which forms a room temperature ionic liquid EMI$^+$-AlCl$_4^-$. The structure yielded by the classical simulations performed within the framework of the polarizable ion model is compared to the results obtained from entirely electronic structure-based simulations: An excellent agreement between the two flavors of molecular dynamics is observed. When changing the organic cation EMI$^+$ by an inorganic cation with a smaller ionic radius (Li$^+$, Na$^+$, K$^+$), the chloroaluminate speciation becomes more complex, with the formation of Al$_2$Cl$_7^-$ in small amounts. The calculated transport properties (diffusion coefficients, electrical conductivity and viscosity) of EMI$^+$-AlCl$_4^-$ are in good agreement with experimental data.
\end{abstract}

\maketitle

\section{Introduction}


When monovalent chloride salts are dissolved in liquid AlCl$_3$, ionic liquids with low melting points are formed. When the added cation is inorganic, the system becomes a molten salt, like NaAlCl$_4$ with $T_m$~=~152~$^\circ$C .\cite{oye2000a} When it is organic, the system is likely to be liquid below 100~$^\circ$C, thus potentially becoming a room temperature ionic liquid (RTIL); for example $T_m$(EMIAlCl$_4$)~=~7~$^\circ$C \cite{fannin1984a} (where EMI$^+$ stands for 1-ethyl-3-methylimidazolium cation).

A number of investigations using molecular dynamics simulations have been performed for ionic liquids directed at either structure or dynamical properties.~\cite{maginn2007a} Given the molecular nature of cations or anions forming ionic liquids, earlier molecular dynamics simulations were performed with pairwise additive potentials based on non-polarizable force fields such as OPLS or AMBER.~\cite{price2001a,deandrade2002a,margulis2002a,morrow2002a,urahata2004a,canongialopes2004a,salanne2006b,siqueira2007a} These simulations have been able to reproduce the structural properties of ionic liquids, for instance showing that a somewhat unusual hydrogen bond can take place in imidazolium based ionic between the imidazolium ring proton and the anion species.~\cite{bhargava2007b} This observation is in agreement with IR and NMR spectroscopy studies.~\cite{wulf2007a,fumino2008a,fumino2008b} At intermediate spatial range (0.5 to 0.8~\AA$^{-1}$), a chain-length dependent pre-peak appears either in the static structure factor calculated from molecular dynamics trajectories~\cite{urahata2006a,annapuredy2010a} or obtained by small wide angle X-ray scattering (SWAXS) and/or small angle neutron scattering (SANS).~\cite{triolo2008a,hardacre2010a,russina2009a}  This feature has been assigned to structural segregation or inhomogeneity, i.e. the imidazolium rings and the alkyl chain regions of cations respectively cluster to form polar and apolar domains in ionic liquids where the cations have long alkyl chains.~\cite{bhargava2007a} Despite their capability to reproduce the structure, the non-polarizable force fields often begin to fail when predicting dynamical properties, such as diffusion coefficient, electrical conductivity, and viscosity.~\cite{lyndenbell2009a} In general the dynamical properties calculated with non-polarizable models are slower than is observed in experiments: The diffusion coefficient and electrical conductivities are lower while the viscosity is greater than the measured values.~\cite{siqueira2007a,cadena2006a,tsuzuki2009a,koddermann2007a,vanoanh2009a} Note that pairwise additive potentials are often parameterised to {\it implicitly} include some aspects of the polarization effects. For example, the simulated fluid may become less viscous if the magnitudes of the partial charges are reduced,~\cite{mullerplathe1995a,bhargava2005a} but sometimes at the price of a poorer representation of other properties. These trends are particularly well-documented in the case of the inorganic systems.~\cite{marrocchelli2010a}

Disregarding the polarization method used, molecular dynamics simulations of molten salts in which polarization effects are included have shown that the liquids become more mobile.~\cite{madden2000a,siqueira2003a} Moreover, molecular dynamics simulations of ionic liquids including polarization have shown better agreement of dynamical properties with experiment, that is, the simulated liquids are either as mobile as the experimental sample or more mobile than the liquids simulated with non-polarizable model.~\cite{yan2004a,schroder2010a,bedrov2010a,yan2010a,yan2010b} Among the studies using a polarizable model for ionic liquids, we would cite the recent and comprehensive work by Yan et al,~\cite{yan2010a,yan2010b} where the authors carefully investigated the role-played by polarization in the ionic liquid EMI$^{+}$ NO$_3$$^{-}$. The authors found that when polarization effects are included, the short-range interactions are enhanced due to the additional charge-dipole and dipole-dipole interactions, while long-range electrostatic interactions become more screened. Accordingly, the imidazolium ring proton - NO$_3^-$ hydrogen bond strength is increased upon switching the polarization on. Another consequence of the stronger polar short-range interactions achieved with the polarizable model for the structure is the enhancement of the tendency towards structural inhomogeneity, which acts by pulling the apolar chains away from the polar center and reinforces the assembly of ethyl chain into the apolar domains. The attenuation of long-range electrostatic interaction is responsible for the increase of diffusion, conductivity and structural relaxation and the reduction of viscosity.

In this work, we study a series of chloroaluminate liquids. A particular emphasis is given to the room temperature ionic liquid consisting of an equimolar mixture of EMICl with AlCl$_3$. We employ two flavors of molecular dynamics simulations, which differ by the method involved for the calculation of the interactions: In the first, they are obtained from a full density functional theory (DFT) calculation, and in the second an analytical force field (including polarization effects) is used. Both methods have already been used successfully for the study of pure AlCl$_3$.~\cite{hutchinson2001a,kirchner2006a} In a first step we detail how the  parameters of the interaction potential for EMICl-AlCl$_3$ have been obtained. The structure and dynamics of this melt are then analyzed and compared to a series of molten salts consisting of equimolar mixtures of AlCl$_3$ with LiCl, NaCl and KCl.

\section{Interaction potentials}
\subsection{Functional form}
In recent work on the simulation of inorganic ionic systems, interaction potentials that were used commonly included many-body effects such as polarization.~\cite{tangney2002a,merlet2010a,salanne2011a,bitrian2011a,giacomazzi2011a} The most generally retained analytical expression~\cite{heaton2006a} is
\begin{eqnarray}
V&=&\sum_{i < j} \left( \frac{q^i q^j}{r^{ij}}+ B^{ij}{\rm e}^{-a^{ij} r^{ij}}  -f^{ij}_6(r^{ij})\frac{C_6^{ij}}{(r^{ij})^6} \right. \\ \nonumber
& & \left.-f^{ij}_8(r^{ij})\frac{C_8^{ij}}{(r^{ij})^8} \right) +V_{\rm pol},
\label{eq:bhm}
\end{eqnarray}
\noindent where the three first terms are pairwise additive. The first term corresponds to the electrostatic interactions, and formal integer charges are used. The two following terms are expressed by a Born-Huggins-Mayer type potential; they consist of an exponentially decaying term accounting for the overlap repulsion of the electronic clouds at short distances, and a term which represents the dispersion effects. The $C_6^{ij}$ and $C_8^{ij}$ are the dipole-dipole and dipole-quadrupole dispersion coefficients; $f_n$ are Tang-Toennies dispersion damping functions~\cite{tang1984a} describing the short-range penetration correction to the asymptotic multipole expansion of dispersion. These functions take the form

\begin{equation}
f_n^{ij}(r^{ij})=1-{\rm e}^{-b_n^{ij}r^{ij}}\sum_{k=0}^n\frac{(b_n^{ij}r^{ij})^k}{k!}
\end{equation}

\noindent where the $b_n^{ij}$ parameter sets the range of the damping effect. The polarization term reflects the distortion of the electronic density in response to the electric fields due to all the other ions. It is written

\begin{eqnarray}
V_{\rm pol}&=&\sum_{i,j}\left[\left(q^i \mu_{\alpha}^j g^{ij}(r^{ij})-q^j \mu_\alpha^i g^{ji}(r^{ij})\right){\mathbb T}_\alpha^{ij}  \right. \\ \nonumber
& & \left.-  \mu_\alpha^i\mu_\beta^j{\mathbb T}_{\alpha \beta}^{ij}\right]  + \sum_i (\frac{1}{2\alpha^i} \mid \bmu^i\mid^2 )
\end{eqnarray}

\noindent where ${\mathbb T}_{\alpha}$ and ${\mathbb T}_{\alpha\beta}$ are the charge-dipole and dipole-dipole interaction tensors and $\alpha^i$ is the polarizability of ion $i$. Again, Tang-Toennies functions are included to account for the short-range damping effects:

\begin{equation}
g^{ij}(r^{ij})=1-c^{ij}{\rm e}^{-b^{ij}r^{ij}}\sum_{k=0}^4\frac{(b^{ij}r^{ij})^k}{k!}
\end{equation}

\noindent These functions differ from the previously defined $f_n^{ij}$ due to the presence of an additional parameter $c^{ij}$, which measures the strength of the ion response to the short-range effects.

The set of induced dipoles $\{\bmu^i\}_{i\in[1,N]}$ is treated as $3N$ additional degrees of freedom of the systems. The dipoles are determined at each time step by minimization of the total polarization energy and they depend on the positions of all the atoms at the corresponding time; therefore the polarization part of the potential is a many-body term.

In the case of room-temperature ionic liquids, molecular species are involved, and additional intramolecular interactions accounting for covalent bonding therefore have to be included. Moreover, the electrostatic interactions now involve {\it partial} charges for each atom and the Lennard-Jones type is more commonly used than the Born-Huggins-Mayer one for the remaining pairwise additive interactions.~\cite{deandrade2002a,margulis2002a,urahata2004a,canongialopes2004a,salanne2006b,siqueira2007a} The repulsion and damping terms are then cast together in the following way:

\begin{equation}
V_{\rm LJ}=\sum_{i<j}4\times \epsilon^{ij}\left[\left(\frac{\sigma^{ij}}{r^{ij}}\right)^{12}-\left(\frac{\sigma^{ij}}{r^{ij}}\right)^{6}\right]
\label{eq:LJ}
\end{equation}

\subsection{Parameterization}

As soon as a functional form has been chosen for the interaction potential, all the atomic and atom pair parameters have to be carefully determined. In all cases, the overall charges on the Cl$^-$, Al$^{3+}$, and EMI$^+$ are fixed as -1, +3, and +1 respectively. Our approach consists in parameterizing the remaining terms on the basis of first-principles electronic structure calculations.~\cite{heaton2006a} This is achieved by a ``force-matching'' method, which is generalized by using the first-principles calculated induced dipoles in addition to the forces in the fitting procedure. Amongst molten chlorides, this method has already been applied successfully to the case of LiCl, NaCl and KCl.~\cite{ohtori2009a} Here we extend this set of parameters in order to include the Al$^{3+}$ and 1-ethyl-3-methylimidazolium cations.

\begin{figure}[h]
\begin{center}
  \includegraphics[width=6cm]{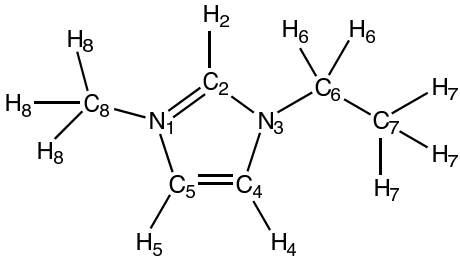}
  \caption{Topology of the 1-ethyl-3-methylimidazolium molecular ion.}
  \label{fgr:topology}
\end{center}
\end{figure}

\begin{table}[h]
  \caption{Interaction potential parameters (partial charges, Lennard-Jones parameters) used for the EMI$^+$ cation.~\cite{canongialopes2004a} Note that the $\epsilon^{i}$ and $\sigma^{i}$ parameters were used for the cation-cation van der Waals interactions only.}
  \label{tbl:parameters3}
  \begin{tabular}{c c c c }
    \hline
    Atom $i$ & $\epsilon^{i}$  & $\sigma^{i}$ & $q^{i}$       \\
             & (kJ mol$^{-1}$) & (\AA)        & ($e$)  \\
    \hline
    N$_1$         & 0.71128   & 3.25     &   0.15  \\
    C$_2$         & 0.29288   & 3.55	    & -0.11  \\
    N$_3$         & 0.71128   & 3.25     &   0.15  \\
    C$_4$         & 0.29288   & 3.55      & -0.13  \\
    C$_5$         & 0.29288   & 3.55      & -0.13  \\
    C$_6$         & 0.27614   & 3.50      & -0.17  \\
    C$_7$         & 0.27614   & 3.50      & -0.05  \\
    C$_8$         & 0.27614   & 3.50      & -0.17  \\
    H$_2$         & 0.12552   & 2.42      &  0.21   \\
    H$_4$         & 0.12552   & 2.42      &  0.21   \\
    H$_5$         & 0.12552   & 2.42      &  0.21   \\
    H$_6$         & 0.12552   & 2.50      &  0.13   \\
    H$_7$         & 0.12552   & 2.50      &  0.06   \\
    H$_8$         & 0.12552   & 2.50      &  0.13   \\
    \hline
  \end{tabular}
\end{table}

In the present work, we have chosen to describe the EMI$^+$ cation as a non-polarizable species for two reasons. Firstly, polarization effects are dominated by the chloride anions (in systems like EMI-BF$_4$ involving less polarizable anions the situation is opposite, with the polarization of the cation playing the main role~\cite{borodin2009a}). Secondly, from a practical point of view, the computer time associated with the minimization procedure of the polarization term increases substantially with the number of polarizable species. For all the intramolecular/intermolecular interactions involving the EMI$^+$ cation only, the parameters were taken from the well-tested force field of P\'adua and co-workers.~\cite{canongialopes2004a} The topology of the molecule is shown on figure \ref{fgr:topology}, and the partial charges and Lennard-Jones individual parameters are given in table \ref{tbl:parameters3}. The pair parameters involved in equation \ref{eq:LJ} are obtained from the usual Lorentz-Berthelot mixing rules:
\begin{eqnarray}
\epsilon^{ij}&=&\sqrt{\epsilon^i \times \epsilon^j} \\
\sigma^{ij}&=&\frac{\sigma^i + \sigma^j}{2}
\end{eqnarray}

\begin{figure}[h]
  \includegraphics[width=\columnwidth]{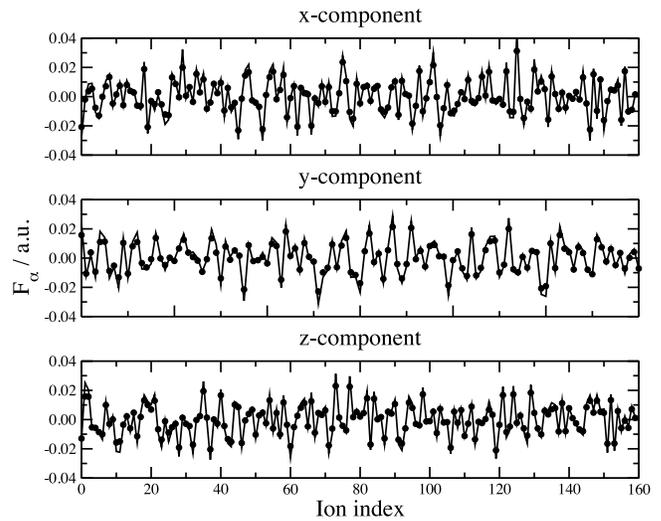}
  \caption{Results of force matching for one EMICl-AlCl$_3$ configuration. The three components of the force acting on each aluminium and chloride ions are compared (dots: DFT, line: PIM (dots: DFT, line: PIM).}
  \label{fgr:fitforces}
\end{figure}

We have then applied the force-matching method to fit all the other interactions, which involve the chloride anion on the one hand and either the aluminium or the EMI$^+$ cation atoms on the other hand. In order to ensure that the force-matching really refines the cation-anion terms, we have held fixed the parameters for the Cl-Cl interactions at the values previously determined for the pure alkali halides.~\cite{ohtori2009a}  To this end, we extracted several configurations from some trajectories obtained from Car-Parrinello molecular dynamics simulations of the liquids EMICl and EMICl-AlCl$_3$. Each of these configurations were then used again as an input to a plane wave density functional theory (DFT) calculation (the details of these calculations are given in the next section). We first performed an energy minimization in order to find the ground-state electronic structure. The forces on each chloride and aluminium (if present) ions were extracted at this stage. Then we transformed the Kohn-Sham orbitals into a set of maximally localized Wannier functions,~\cite{marzari1997a} from which we could determine the induced dipole moment on the Cl$^-$ anions. The parameters in the polarizable potentials are then optimized by matching the dipole and forces from the potential to the DFT calculated ones.~\cite{heaton2006a,aguado2003b} Two distinct functions are minimized. The first one corresponds to the polarization part:

\begin{equation}
\chi_D^2=\frac{1}{N_{\rm Cl^-}}\sum_{i=1}^{N_{\rm Cl^-}}\frac{\mid\bmu^i_{\rm DFT}-\bmu^i_{\rm PIM}\mid^2}{\mid\bmu^i_{\rm DFT}\mid^2}
\end{equation}

\noindent where the subscript PIM stands for ``Polarizable Ion Model''. Since the chloride anion is the only polarizable species, for a given atom type $\alpha$ the two short-range damping parameters $b^{\rm Cl \alpha}$ and $c^{\rm Cl \alpha}$ were optimized with this function. The second function,

\begin{eqnarray}
\chi_F^2&=&\frac{1}{N_{\rm Cl^-}}\sum_{i=1}^{N_{\rm Cl^-}}\frac{\mid{\bf F}^i_{\rm DFT}-{\bf F}^i_{\rm PIM}\mid^2}{\mid{\bf F}^i_{\rm DFT}\mid^2} \\ \nonumber
& & +\frac{1}{N_{\rm Al^{3+}}}\sum_{i=1}^{N_{\rm Al^{3+}}}\frac{\mid{\bf F}^i_{\rm DFT}-{\bf F}^i_{\rm PIM}\mid^2}{\mid{\bf F}^i_{\rm DFT}\mid^2},
\end{eqnarray}

\noindent is used to parameterize the repulsion term of the interaction potential, i.e. for determining the various $B^{ij}$ and $a^{ij}$ values in equation \ref{eq:bhm}.  A well known limitation of the PBE functional that we used in our DFT calculation is that it does not take into account properly the dispersion interactions,~\cite{zahn2008b} so that the $C^{ij}_6$ and $C^{ij}_8$ cannot be fitted by this procedure. Instead, we have used our previously determined Cl$^-$-Cl$^-$ dispersion coefficients~\cite{ohtori2009a} for calculating the various $C^{ij}_n$ ($n$~=~6,8) by using the Slater-Kirkwood relationship.

The quality of the representation of the forces is illustrated in figure \ref{fgr:fitforces} where we compare the three components of the forces acting on the aluminium (32 first atoms) and chloride ions calculated with our potential with the DFT calculated data, for one of the EMICl-AlCl$_3$ configurations. The agreement is seen to be uniformly good. For both RTIL systems, we obtained values of 0.13 for $\chi^2_F$, which is similar to our previous work on molten LiCl, NaCl and KCl. Concerning the dipoles, the values of $\chi_D^2$ are respectively 0.39 and 0.03 on average for the EMICl and EMICl-AlCl$_3$ configurations. The poorer agreement in the first case is due to the fact that the electric field felt by the Cl$^-$ anions, which is only due to the EMI$^+$ cation, is only partially reproduced by the partial charges of the atoms. When aluminium cations are added to the melt, this electric field becomes dominated by the trivalent charges they carry, and this effect being much more easily captured by our model. The observation suggests that further refinement of the charges of the EMI$^+$ model would be a good target to further improve the predictive power of the simulations.

\begin{table}[h]
  \caption{Parameters for the pairwise additive terms of the interaction potential. $b_6^{ij}$~=~$b_8^{ij}$~=~3.21~\AA$^{-1}$ for all pairs.}
  \label{tbl:parameters1}
  \begin{tabular}{c c c c c }
    \hline
    Atom pair & $B^{ij}$ $\times 10^{-6}$       & $a^{ij}$     & $C_6^{ij}$ $\times 10^{-6}$              & $C_8^{ij}$ $\times 10^{-6}$              \\
       $i-j$  & (J mol$^{-1}$) & (\AA$^{-1}$) & (J mol$^{-1}$ \AA$^6$) & (J mol$^{-1}$ \AA$^8$)  \\
    \hline
    Cl-Cl     & 722.0 & 3.396 &  8.07 & 4.52    \\
    Cl-Al     & 95.8 & 2.955 & 0.0 & 0.0 \\
    Cl-Li   & 154.5 & 3.685 & 0.0 & 0.0  \\
    Cl-Na   & 177.2 & 3.262 & 2.16 & 2.70  \\
    Cl-K    & 547.5 & 3.458 & 1.52 & 1.91  \\
    Cl-C$_i$ & 331.8 & 3.376 & 2.39 & 1.34  \\
    Cl-N$_i$ & 260.4 & 4.401 & 1.44 & 0.81  \\
    Cl-H$_i$ &  10.8 & 2.901 & 0.41 & 0.23 \\
    \hline
  \end{tabular}
\end{table}

The parameters which were determined from the fitting procedure are summarized in tables \ref{tbl:parameters1} and \ref{tbl:parameters2}. No distinction has been made between the various H$_i$, C$_i$ and N$_i$ atom types because it did not lead to any improvement of the fit quality.

\begin{table}[h]
  \caption{Parameters for the polarization part of the interaction potential. The polarizability of the Cl$^-$, Na$^+$ and K$^+$  ions were respectively set to 2.96, 0.13 and 0.70~\AA$^3$. }
  \label{tbl:parameters2}
  \begin{tabular}{c  c c c}
    \hline
    Atom pair &   $b^{ij}$   & $c^{ij}$   &  $c^{ji}$ \\
       $i-j$  &   (\AA$^{-1}$) & (--) &  (--) \\
    \hline
    Cl-Al   &  3.802 & 2.953 & --\\
    Cl-Li   &  3.517 & 2.079  & -- \\
    Cl-Na   & 3.326  & 3.000  & 0.697\\
    Cl-K    &  3.084 & 3.000  & 0.917\\
    Cl-H$_i$ &   2.901 & 0.301 & -- \\
    \hline
  \end{tabular}
\end{table}

\section{Simulation details}
\subsection{Electronic structure-based molecular dynamics}

For the ab initio-molecular dynamics simulations we used density functional
theory with the generalised gradient approximation of Perdew, Burke and
Ernzerhof, PBE\cite{perdew1996a} as the exchange-correlation term in the Kohn-Sham
equations, and we replaced the action of the core electrons on the valence
orbitals with norm-conserving pseudo potentials of the Troullier-Martins type.~\cite{troullier1991a,troullier1991b} We expanded the wave functions in plane waves up to the cut-off energy
of 70~Ry. We sampled the Brillouin zone at the $\Gamma$ point, employing
periodic boundary conditions.

We performed the simulations in the NVT ensemble, employing a Nos\'{e}-Hoover
thermostat at a target temperature of 300~K and a characteristic frequency of
595 cm$^{-1}$ , a stretching mode of the AlCl$_3$ molecules. We propagated the
velocity Verlet equations of motion with a time step of 5~a.t.u. = 0.121~fs,
and the fictitious mass in the Car-Parrinello molecular dynamics~\cite{car1985a} for
the electrons was chosen as 700~a.u.; we note that the fictitious dynamics
increases the temperature seen by the ions somewhat.~\cite{tangney2006a} A cubic
simulation cell with a edge length of 22.577~\AA{} containing both 32 anionic
and cationic molecules, equalling to the experimental density of
1.293~g/cm$^3$. The length of the trajectory was 20~ps. The CPMD simulation code was used.~\cite{cpmd}

\subsection{Polarizable ion model-based molecular dynamics}
For the inorganic molten salts (LiCl-AlCl$_3$, NaCl-AlCl$_3$ and KCl-AlCl$_3$) the simulations were performed in the NVT ensemble, employing a Nos\'e-Hoover thermostat at a target temperature of 573~K, with a time step of 0.5~fs, for a total simulation time of 3~ns. For EMICl-AlCl$_3$, several simulations were performed at the temperatures of 298, 433, 473, 523 and 573~K, with a time step of 1~fs. Note that the higher temperatures employed may be above the thermal decomposition temperature of the ionic liquid (no data could be found in the litterature); these simulations were performed in order to make a fair comparison of the structure and dynamics with the inorganic molten salts.  The total simulation time was of 1.5~ns for each temperature except 298~K, for which it was of 3~ns. The FIST module of CP2K simulation package was used.~\cite{cp2k} In all cases, cubic simulation cells were employed. The total number of atoms and the cell edge length, which were chosen according to the experimental densities (an extrapolation of the room temperature data was made for EMICl-AlCl$_3$ when necessary),~\cite{janz1988a,fannin1984a}, are provided in table \ref{tbl:simcond}. The cut-off distance for the real space of the Ewald sum and the short range potential was set to 13~\AA\ for  EMICl-AlCl$_3$ and to half the size of the simulation cell otherwise. 

\begin{table}[h]
  \caption{Number of atoms and cell edge length ($L$) employed in the simulations ($M^+$~=~Li$^+$, Na$^+$, K$^+$ or EMI$^+$).}
  \label{tbl:simcond}
  \begin{tabular}{l c c c c c}
    \hline
System & T (K) & $ N_{\rm M^+} $ &  $N_{{\rm Al}^{3+}}$ & $ N_{\rm Cl^-} $ & $L$ (\AA) \\
    \hline
LiCl-AlCl$_3$ & 573 & 120 & 120 & 480 & 28.426 \\
    \hline
NaCl-AlCl$_3$ & 573 & 120 & 120 & 480 & 28.790 \\
    \hline
KCl-AlCl$_3$ & 573 & 120 & 120 & 480 & 29.728 \\
    \hline
EMICl-AlCl$_3$ & 298 & 200 & 200 & 800 & 41.576 \\
               & 433 & 200 & 200 & 800 & 42.807 \\
               & 473 & 200 & 200 & 800 & 43.200 \\
               & 523 & 200 & 200 & 800 & 43.713 \\
               & 573 & 200 & 200 & 800 & 44.251 \\
    \hline
  \end{tabular}
\end{table}

\section{Results and discussion}

\subsection{Chloroaluminate speciation}

 Unlike previous classical MD simulations of chloroaluminate-based ionic liquids, in which the AlCl$_4^-$ ion was treated as a molecular species,~\cite{deandrade2002a,deandrade2002b} an advantage of the two simulation methods involved in the present study is that they both treat the chloride and aluminium ions as independent species. This is done intrinsically in electronic structure based simulations on one hand, and by construction of the polarizable ion model (PIM) on the other. This capability has been used, for example, in pure AlCl$_3$ liquid, in which structure is characterized by the formation of AlCl$_4^-$ tetrahedra sharing an edge to form Al$_2$Cl$_6$ dimers  or larger clusters. MD simulation studies either employing similar interaction potentials as in the present study~\cite{hutchinson1999b} or based on electronic structure calculations~\cite{kirchner2006a,east2007a} were able to confirm quantitatively this picture, in agreement with X-ray diffraction experiments.~\cite{harris1951a} The formation of large chloroaluminate species was also investigated in simulations of a system consisiting of one EMICl pair dissolved in a AlCl$_3$ solvent.~\cite{kirchner2007a}

In equimolar mixtures of AlCl$_3$ with monovalent chloride $M$Cl salts, the main ``chemical reaction'' involving chloroaluminate species which is expected to occur\cite{fannin1984b,lipsztajn1985a,hussey1986a} is:

\begin{equation}
{\rm AlCl}_4^- + {\rm AlCl}_4^- \rightarrow {\rm Al}_2{\rm Cl}_7^- + {\rm Cl}^-
\label{eq:chemicalreaction}
\end{equation}

Here in the case of the equimolar EMICl-AlCl$_3$ the two simulation methods agree on the formation of isolated AlCl$_4^-$ anions only, in agreement with experimental results.~\cite{lipsztajn1985a} Small differences are observed in the first neighbour distances, which are shorter in the simulation employing the polarizable model: The first peaks of the Al-Cl radial distribution functions respectively appear at 2.18~\AA\ and 2.22~\AA\ for PIM-MD and CPMD, while the corresponding Cl-Cl functions display some first peak maxima at 3.53~\AA\ and 3.61~\AA. These differences may be attributed to the choice of keeping the Cl$^-$-Cl$^-$ interaction potential parameters similar to the values which had been previously been determined for pure LiCl, NaCl and KCl molten salts.~\cite{ohtori2009a} Another source of difference could be the presence of attractive dispersion effects in the classical MD only.

\begin{figure}[h]
  \includegraphics[width=\columnwidth]{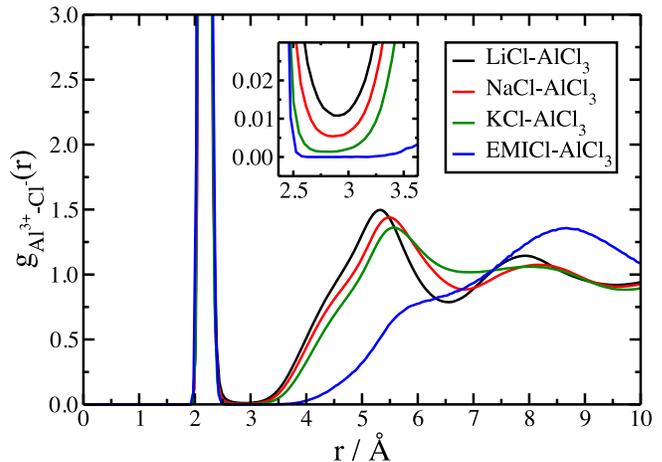}
  \caption{Al$^{3+}$-Cl$^-$ radial distribution functions for the $M$Cl-AlCl$_3$ mixtures at 573~K. Insert: Enlargment of the first minimum zone.}
  \label{fgr:rdfalcl}
\end{figure}

When $M^+$ is an inorganic cation (Li$^+$, Na$^+$, K$^+$), we observe a weakening of the Al-Cl links. This may be quantified from the corresponding radial distribution functions (RDF), for which we observe a decrease of the intensity of the first peak, which switches from  a value of 47 in EMICl-AlCl$_3$ to one of 18 in LiCl-AlCl$_3$ (not shown). Some Cl$^-$ ions begin to jump from one aluminium coordination sphere to another. The signature of these exchanges can also be tracked in the RDFs which are given on figure  \ref{fgr:rdfalcl}: The minimum occurring after the first maximum no longer corresponds to a value of zero as in the case in EMICl-AlCl$_3$. The value of this minimum also increases when the ionic radius of the $M^+$ cation decreases in the series. The position of the second peak of the RDF is slightly shifted toward larger distances when passing from Li$^+$ to Na$^+$ and then K$^+$, but this distance increases by as much as $\approx$~3\AA\ when passing from K$^+$ to EMI$^+$ (note that the peak also becomes much wider, with a pronounced shoulder). The absence of exchange of Cl$^-$ between AlCl$_4^-$ coordination tetrahedra in EMICl-AlCl$_3$ can therefore easily be explained by the longer distance that has to be crossed during a jump.

\begin{figure}[h]
  \includegraphics[width=\columnwidth]{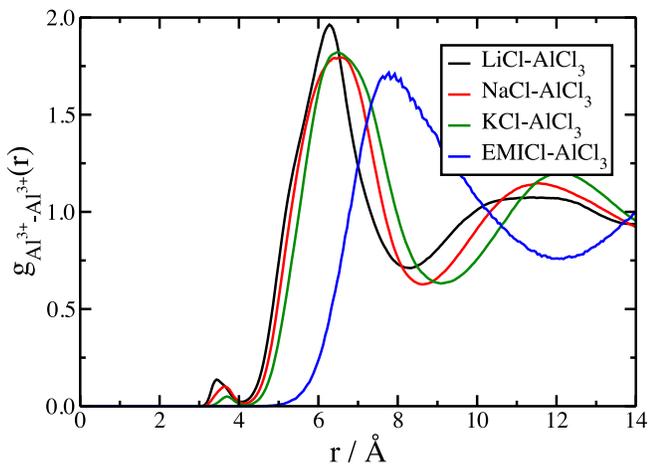}
  \caption{Al$^{3+}$-Al$^{3+}$ radial distribution functions for the $M$Cl-AlCl$_3$ mixtures at 573~K.}
  \label{fgr:rdfalal}
\end{figure}

Going back to the chemical reaction defined in equation \ref{eq:chemicalreaction}, we observe some signature of the formation of Al$_2$Cl$_7^-$ ions in our systems by examining the Al-Al RDFs (shown on figure \ref{fgr:rdfalal}). A first peak with a small intensity is obtained in all the systems except EMICl-AlCl$_3$. The corresponding distance (ranging from 3.5\AA\ in LiCl-AlCl$_3$ to 3.7~\AA\ in KCl-AlCl$_3$) is smaller than the length of two Al-Cl bonds, indicating pairs of Al$^{3+}$ ions linked by common Cl$^-$ anions.\cite{hutchinson1999b} To quantify the proportion of each chloroaluminate species present in the melt, we have followed the same procedure as in our previous work on LiF-BeF$_2$ mixtures, where we could provide a quantitative picture of the speciation in the melt for a wide range of composition.~\cite{salanne2006a,salanne2007b} Here we have chosen to define an Al-Cl-Al bond when two Al-Cl and the corresponding Al-Al distances are shorter than the corresponding RDFs minima. By extending the analysis of the linkages between ions, we can calculate the proportion of any different polynuclear ionic species present in the melt. The main species that we observed apart from the isolated AlCl$_4^-$ anions were the Al$_2$Cl$_7^-$ and Al$_3$Cl$_{10}^-$ species; their proportions are provided in table \ref{tbl:chloroaluminates} for a temperature of 573~K. Note that these numbers have to be taken cautiously: Due to the finite size of the simulation cells and to the relatively low number of chloride anions jumping or bridging events, some refinement may be necessary. We also observed some small proportions of transient chemical species such as AlCl$_5^{2-}$ or Al$_2$Cl$_8^{2-}$ in negligible amounts.
\begin{table}[h]
  \caption{Proportions (given in percentage of aluminium ions) of the various chloroaluminate species at 573~K.}
  \label{tbl:chloroaluminates}
  \begin{tabular}{l  c c c }
    \hline
  System & [AlCl$_4^-$] & [Al$_2$Cl$_7^-$] & [Al$_3$Cl$_{10}^-$] \\
    \hline
LiCl-AlCl$_3$ & 94.1 & 3.5 & 0.7    \\
NaCl-AlCl$_3$ & 95.7 & 3.4 & 0.2   \\
KCl-AlCl$_3$  & 98.2 & 1.6 & 0.0    \\
EMICl-AlCl$_3$& 100.0 & 0.0 & 0.0   \\
    \hline
  \end{tabular}
\end{table}

\subsection{Effect of the $M^+$ cation nature on the intermolecular structure}
\begin{figure}[h]
  \includegraphics[width=\columnwidth]{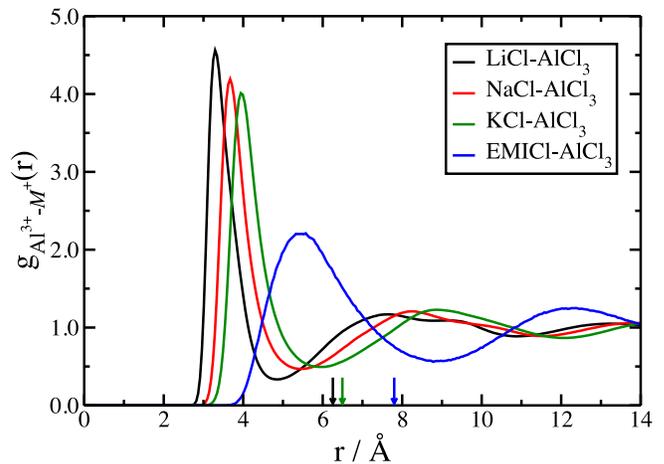}
  \caption{AlCl$^-_4$-$M^+$ radial distribution functions at 573~K. In the case where $M^+$ is the EMI$^+$ cation, the position of the center of mass of the molecule was used to calculate the function. Small arrows indicate the position of the corresponding Al$^{3+}$-Al$^{3+}$ RDF first maximum (occulting the first peak due to the formation of Al$_2$Cl$_7^-$ ions); the Na$^+$ one is not visible because it is confounded with the K$^+$ one.}
  \label{fgr:rdfcatan}
\end{figure}

From the analysis of speciation in the chloroaluminate species we can try to describe the equimolar mixtures of $M$Cl with AlCl$_3$ as generic $MX$ molten salts (like e.g. alkali halides), where $M^+$ is the monovalent cation and $X^-$ is the AlCl$_4^-$ anion. In simple $MX$ molten salts, the structure around a given ion consists in alternating layers of opposite charges. As a result, the RDFs are characterized by strong oscillations, with the like-like RDFs maxima (minima) which are located at the same positions as the cation-anion RDF minimum (maximum). In the present work, we can compare the  $M^+$-Al$^{3+}$ RDF to the Al$^{3+}$-Al$^{3+}$ one; the Al$^{3+}$ was chosen because it gives the AlCl$_4^-$ ion center of mass; for the EMI$^+$ cation the $M^+$-Al$^{3+}$ was calculated with respect to the imidazolium ion center of mass position. These functions are shown on figures \ref{fgr:rdfalal} and \ref{fgr:rdfcatan}. On the latter we also report the position of the Al$^{3+}$-Al$^{3+}$ RDF first maximum (omitting the small peak due to the formation of Al$_2$Cl$_7^-$) with an arrow on the abcissa in order to facilitate the comparison with the position of the first minimum in the cation-anion term. We observe that these arrows are not located at the same position as the corresponding minima, but the difference remains small. This means that the structure deviates slightly from that of a simple $MX$ molten salt. This conclusion still holds for EMICl-AlCl$_3$ at lower temperatures: The main features of the $M^+$-Al$^{3+}$ RDF remain unchanged when passing from 573~K to 298~K.

 From this figure we can observe somewhat different behaviour depending on the size of the $M^+$ cation. In the case of the LiCl-AlCl$_3$ melt, the $M^+$-Al$^{3+}$ RDF minimum is located at a shorter distance than the corresponding Al$^{3+}$-Al$^{3+}$ RDF first maximum; this is due to the small size of the Li$^+$ cation, which allows it to ``penetrate'' slightly into the AlCl$_4^-$ anion van der Waals sphere. The difference becomes smaller for the NaCl-AlCl$_3$ and KCl-AlCl$_3$ melts for which the extrema of the two functions almost coincide, and the situation is reversed for EMICl-AlCl$_3$ where closer packing of the anions is allowed. The $M^+$-Al$^{3+}$ RDF first peak is also broader in the case of EMICl-AlCl$_3$ compared to the alkali cations containing systems, but the difference is not dramatic. The two latter observations suggest that in the present system the EMI$^+$ cation behaves mostly like a monoatomic cation with a very large ionic radius. In the next section we will investigate its coordination structure on a more local scale, in order to understand if the charge delocalization as well as the presence of different atom types induce a particular ordering in the first solvation shell.

\subsection{EMICl-AlCl$_3$ equimolar mixture}

\begin{figure}[h]
  \includegraphics[width=\columnwidth]{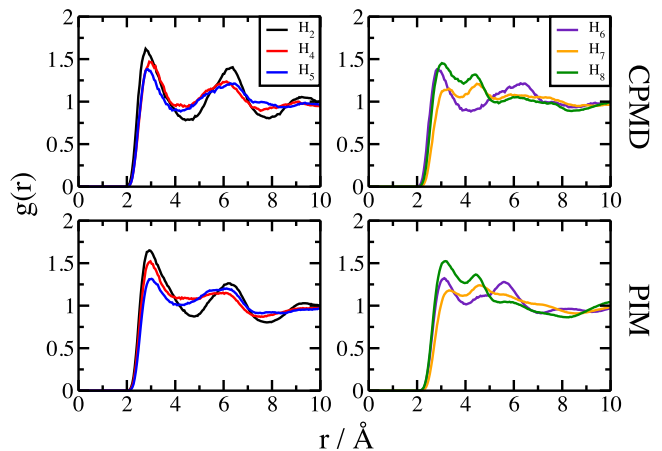}
  \caption{Radial distribution functions of the chloride anions and the protons from the EMI$^+$ cations. Top: CPMD results; bottom: PIM results. Left: Protons from the imidazolium ring. Right: Protons from the alkyl chains.}
  \label{fgr:rdfrtil}
\end{figure}

\begin{figure}[h]
  \includegraphics[width=\columnwidth]{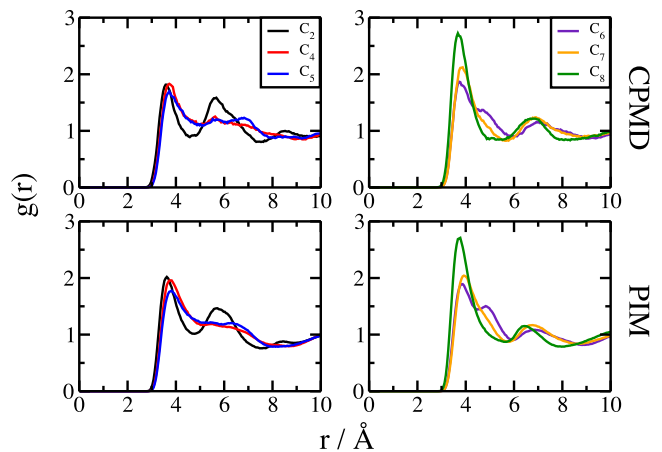}
  \caption{Radial distribution functions of the chloride anions and the carbon atoms from the EMI$^+$ cations. Top: CPMD results; bottom: PIM results. Left: Carbon atoms from the imidazolium ring. Right: Carbon atoms from the alkyl chains.}
  \label{fgr:rdfrtilC}
\end{figure}

In order to characterize further the intermolecular structure of the EMICl-AlCl$_3$ system, the radial distribution functions of the chloride anions and the atoms from the EMI$^+$ cations have been calculated. They are  plotted on figures \ref{fgr:rdfrtil} (hydrogen atoms) and \ref{fgr:rdfrtilC} (carbon atoms). The CPMD and PIM results obtained at room temperature are compared; an excellent agreement is observed for all the atoms. The small differences observed are more likely due to the poorer sampling in the CPMD simulation rather than to a deficiency of the classical interaction potential. In a previous attempt to obtain classical force field parameters through a force-matching procedure similar to ours on the ionic liquid dimethylimidazolium chloride (DMICl), Youngs {\it et al.} have been less successful in reproducing the structure obtained from CPMD.~\cite{youngs2006a} Although they were able to reproduce the peak positions better  than with previous classical MD potentials, the intensities of the first peaks where largely overestimated (especially in the case of the ring protons). The over-structuring obtained by these authors could partly be cancelled by multiplying all the $\epsilon^i$ parameters by a factor of 2, at the price of a worse reproduction of the initial set of DFT forces (therefore leading to higher $\chi^2_F$ values). The better agreement observed here can be attributed to two factors. The main one is the use of a more complicated functional form for the analytical interaction potential, or in other words to the inclusion of anion polarization effects. The second is the fact that we focused on the chloride (and aluminium) ion forces in our force-fitting process; in Youngs {\it et al.} study all the parameters, including the intramolecular ones, were fitted during the procedure. The intramolecular contribution may have somewhat ``masked'' the intermolecular ones, thus leading to a smaller precision for the latter. It is worth noticing that the better agreement observed here was not obtained by fitting more parameters. Due to the choice of using the set of partial charges of Padua and co-workers for the EMI$^+$ ion atoms, only twelve parameters involving the Cl$^-$ and EMI$^+$ species had to be fitted in the parameterization process: The eight Cl-C$_i$,N$_i$,H$_i$,Al repulsion parameters ($B^{ij}$ and $a^{ij}$) and the four parameters $b^{ij}$ and $c^{ij}$ involved in the damping of the chloride ions induced dipoles by the hydrogen and aluminium atoms.

\begin{figure}[h]
  \includegraphics[width=4.0cm]{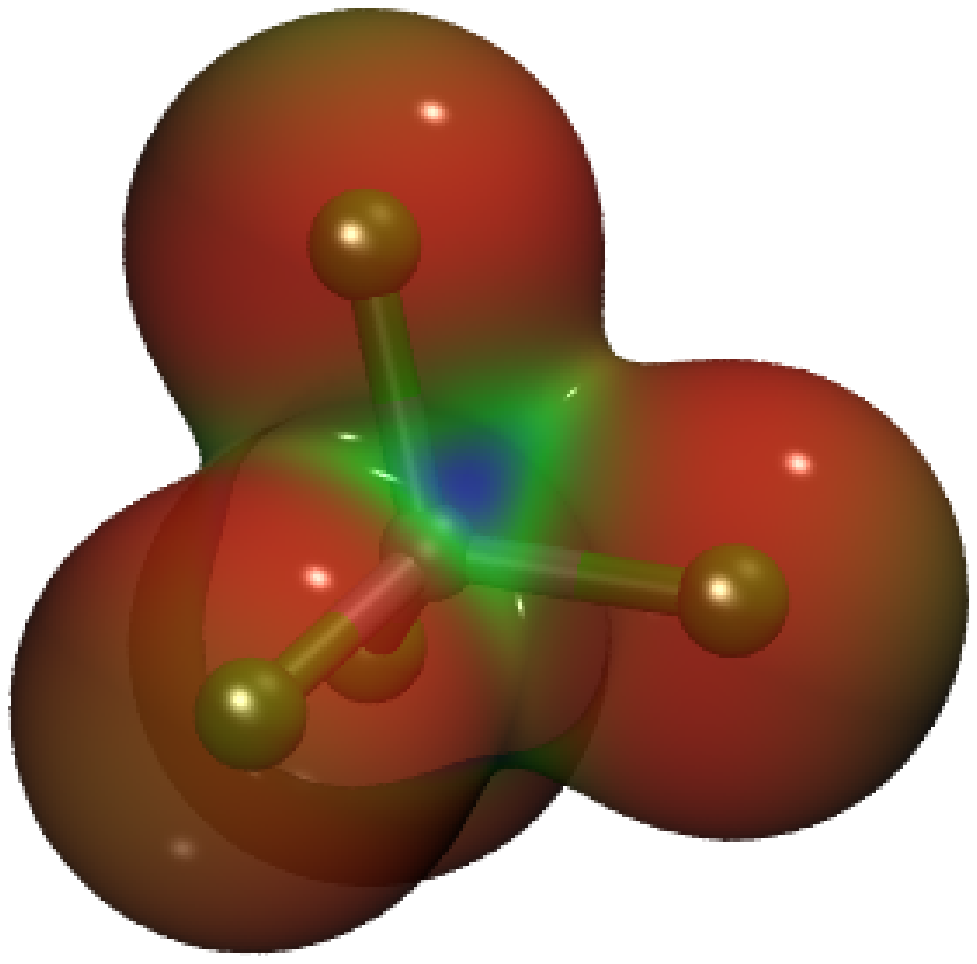}  \includegraphics[width=7.0cm]{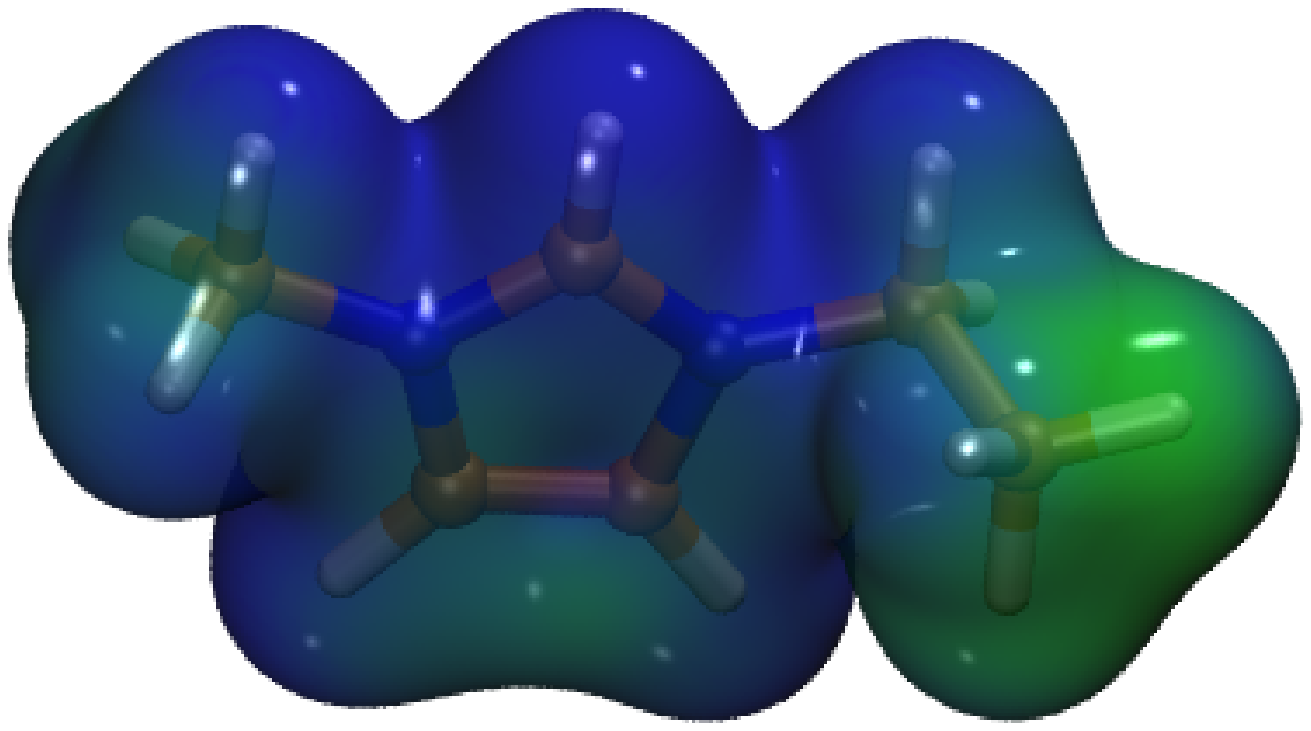}

  \caption{The electrostatic potential mapped onto the electron density with a surface values of 0.067~e$^-$.\AA$^{-3}$. The colour scale ranges from -0.1 (red) to +0.1 (blue) atomic units. Left: AlCl$_4^-$ , right: EMI$^+$.}
  \label{fgr:electrostatic}
\end{figure}

A striking feature of the radial distribution functions involving the chloride ions and the EMI$^+$ atoms is that the most intense peaks are observed for the carbon atoms. The peak is more pronounced for the C$_7$ and C$_8$, while its intensity is almost the same for the other carbons.  This result contrasts sharply with some other RTILs, like the above-mentioned DMICl for example, for which the anions tend to stay in the vicinity of the imidazolium ring protons.~\cite{delpopolo2005a,youngs2006a} On the contrary, it resembles to the case of the 1-$n$-butyl-3-methylimidazolium hexafluorophosphate, for which the hydrogen bonds were found to be weaker than expected, as indicated by their short lifetimes.~\cite{zhao2009a}  In the present system the situation is due to the huge electrostatic interaction between Cl$^-$ anions and the (Lewis) acidic Al$^{3+}$ cation. To illustrate this the electrostatic potential mapped onto the isosurface of the electron density of the two individual ions is shown in figure \ref{fgr:electrostatic}. It provides insight into the charge distribution of each ion. For the AlCl$_4^-$ we recognize in the red color (low electrostatic potential) the consequence of the negative charge that is associated with this ion. The negative charge is distributed all over the chlorine atoms, with a higher negative charge on the inner side (the sphere-like surface around the chloride is coloured in dark red on the interior and in lighter red at the exterior of the AlCl$_4^-$ ion). Towards the center, close to the aluminium atom, we find a decreased negative charge shown as the blue color in the left panel of Fig. 6. The opposite is the case for the EMI$^+$. Here the positive charge leads to the blue color (high electrostatic potential) around this ion. Upon closer inspection we find that the ring protons show the same blue color as most of the molecule. A slight decrease of the charges can be found in the methyl group protons and a stronger decrease (green color) can be found at the terminal ethyl protons. This is in accordance with the observation from the radial pair distribution functions and with chemical intuition that these protons are less acidic than the other protons of the cation.

In the PIM simulations, the chloride anions therefore have their induced dipoles oriented along the Al-Cl axis, which hinders the formation of hydrogen bonds. If we look in further detail at the chloride-proton radial distribution functions, we see that the H$_2$-Cl$^-$ one shows the most pronounced peak, associated to the shortest separation. It is therefore clear that this proton is the most acidic coordination site. The functions involving the other two protons of the imidazolium ring (H$_4$ and H$_5$) also display some peaks at slightly larger distances. However, considering the protons from the ethyl and the methyl group, small pronounced peaks can also be observed. The methyl group hydrogen (H$_8$) function first maximum has the same height as the H$_4$ one; a similar observation is made for the terminal hydrogen atoms of the ethyl group (H$_6$) compared to the ring H$_5$ protons.

\begin{figure}[h]
  \includegraphics[width=5.5cm]{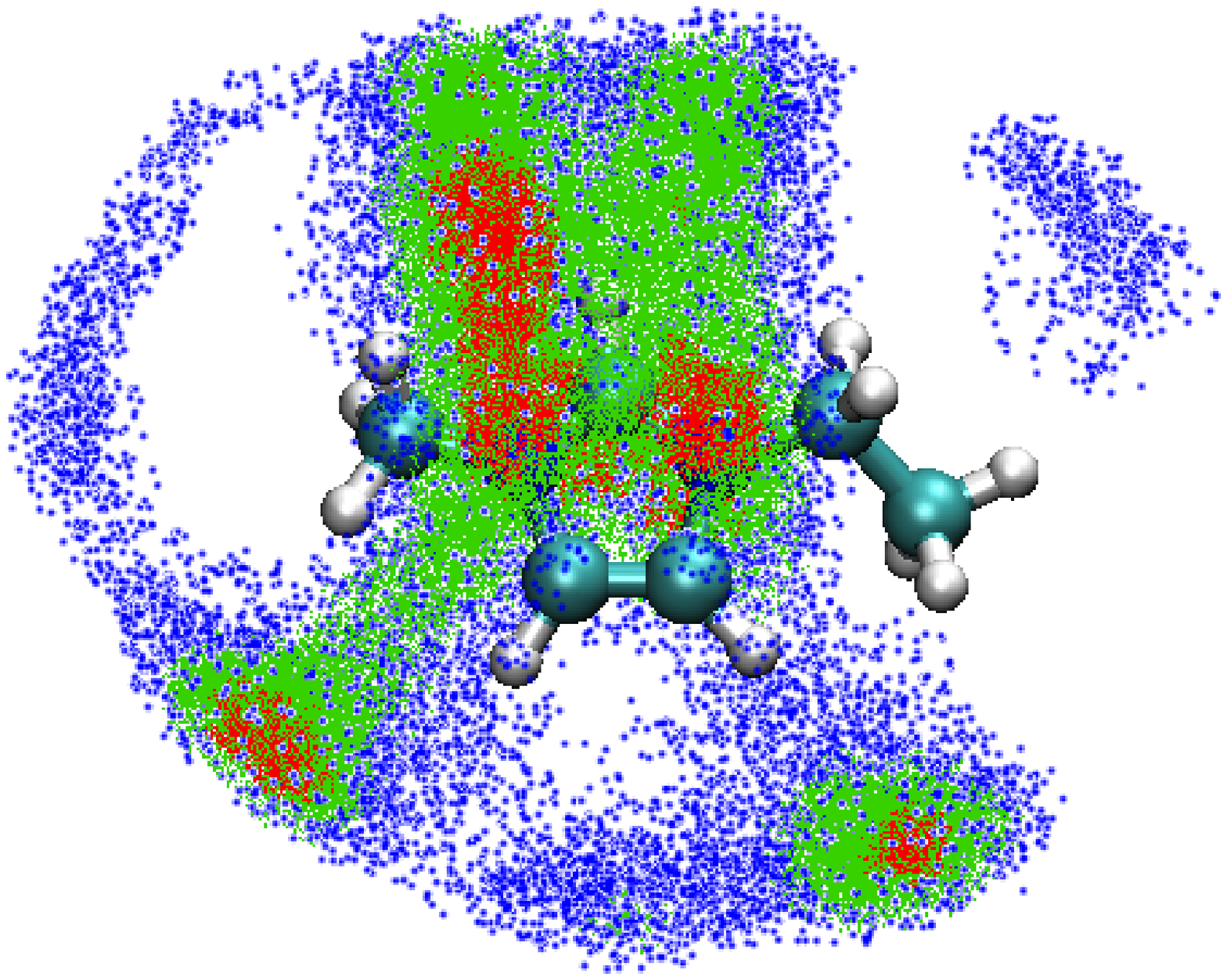} \includegraphics[width=5.5cm]{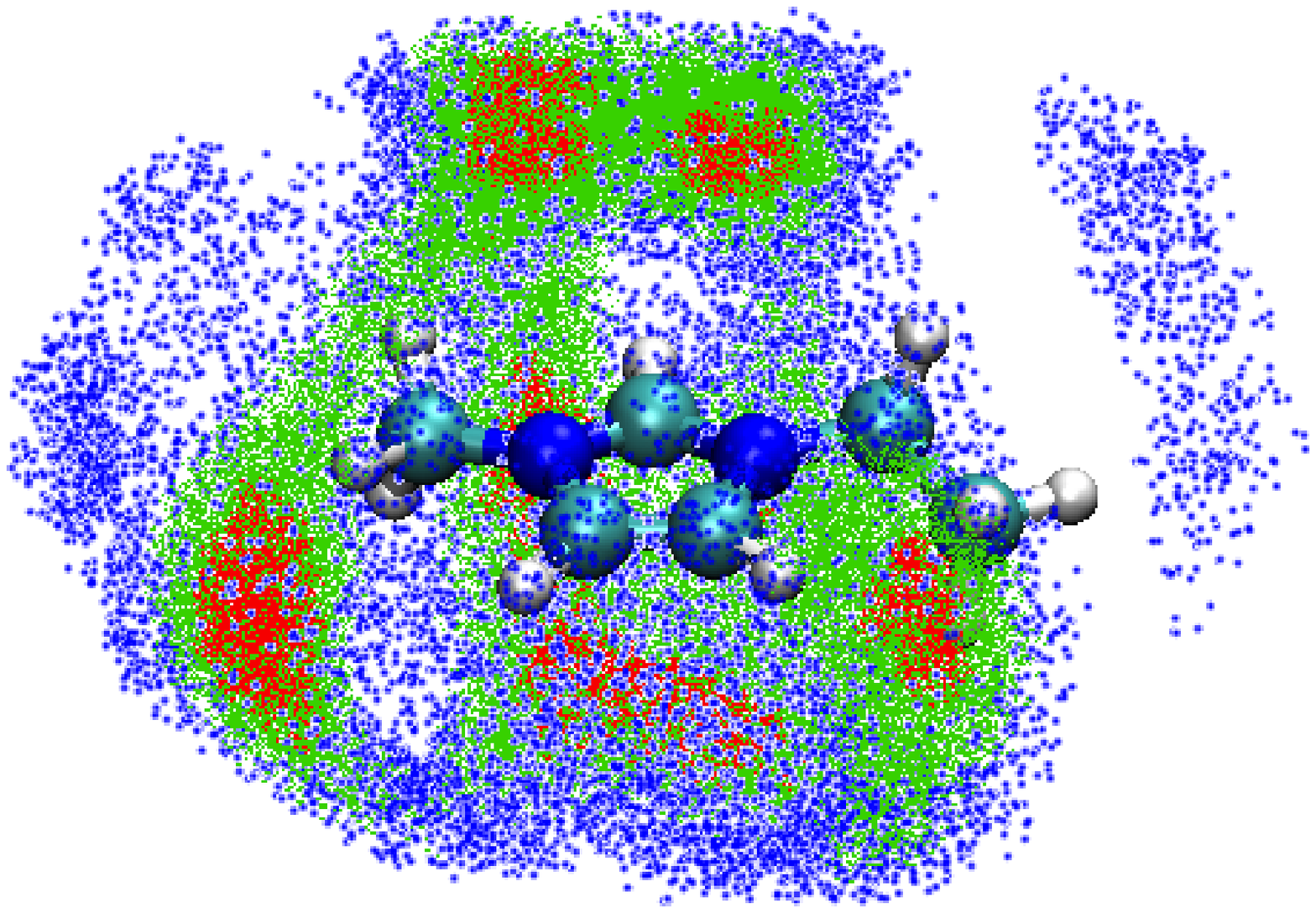}
  \caption{Probability density maps of location of nearest-neighbor anions around the imidazolium ring in equimolar EMICl-AlCl$_3$ mixture at 298~K. The left panel shows the view above the imidazolium ring plane, and the bottom panel shows the view on the ring plane. Red, green and blue regions correspond respectively to high, intermediate and low probability of finding an anion around a given cation. By high probability we mean higher than 80~\% of the maximum density of occurences, and low probability means in between 40~\% and 60~\% of the maximum density of occurences (probability lower than 40~\% is not shown).}
  \label{fgr:maps}
\end{figure}

The arrangement of the chloride anions around the imidazolium cations therefore seems to correspond more to an involved network including the alkyl chains as well as the ring of the imidazolium cation, rather than to individual ion pairs, confirming the results obtained for many imidazolium-based ionic liquids.~\cite{zhao2009a,kohagen2011a} In order to confirm this picture, the three-dimensional density map was calculated for the chloride anions located in the first solvation shell of the EMI$^+$. Views of these maps perpendicular and parallel to the imidazolium ring are provided on figure \ref{fgr:maps}. Red, green and blue regions correspond respectively to high, intermediate and low probability of finding an anion around a given cation. By high probability we mean higher than 80~\% of the maximum density of occurences in these figures, and low probability means in between 40~\% and 60~\% of the maximum density of occurences (probability lower than 40~\% is not shown). The important features of the radial distribution functions, i.e. a preference for some regions close to the C$_7$ and C$_8$ carbons and an absence of important correlations with the protons, are recovered. In general, localization effects are much less important than in the DMICl case, and the formation of hydrogen bond is not the main influence governing the structure of EMICl-AlCl$_3$. It is worth mentioning that in a study involving imidazolium cations with a series of halide anions of different size, the analysis of density maps showed that the larger the anion the lower is the probability of the existence of hydrogen bond between the anion and the cation protons.~\cite{urahata2004a} The present results agree with this view when considering that the relevant anion in our system is the full AlCl$_4^-$ entity.  

The structural analysis of the equimolar EMICl-AlCl$_3$ mixture shows that it can almost be viewed as a classical $MX$ molten salt, where $M^+$~=~EMI$^+$ and $X^-$~=~AlCl$_4^-$, with a structure characterized by a competition between close packing and charge ordering effects occuring between the two species. Small deviations from the $MX$ system are observed due to the existence of weak spatial correlations arising from the formation of hydrogen bonds between the ring protons and the chloride anions.

\subsection{Transport}

In the recent simulation work on the room temperature ionic liquids, it was shown that polarization effects are important for reproducing accurately the transport properties.~\cite{yan2004a,schroder2010a,bedrov2010a,yan2010b,borodin2009a} Here from the Einstein relation
\begin{equation}
D_\alpha = \lim_{t\rightarrow \infty}\frac{1}{6t} \langle \mid \boldsymbol{\delta}{\bf r}_i(t)\mid^2\rangle
\end{equation}

\noindent where $\boldsymbol{\delta}{\bf r}_i(t)$ is the displacement of a typical ions of species $\alpha$ in time $t$, we obtained a diffusion coefficient of 1.04$\times$10$^{-10}$~m$^2$ s$^{-1}$ for the cation at 298~K. This compares very well with the experimental value measured by pulse-field gradient NMR spectroscopy, which is 0.95$\times$10$^{-10}$~m$^2$ s$^{-1}$ at 293~K.~\cite{larive1995a} For the anion we obtained a diffusion coefficient of 4.45$\times$10$^{-10}$~m$^2$ s$^{-1}$ at 298~K.

\begin{figure}[h]
\begin{center}
  \includegraphics[width=.8\columnwidth]{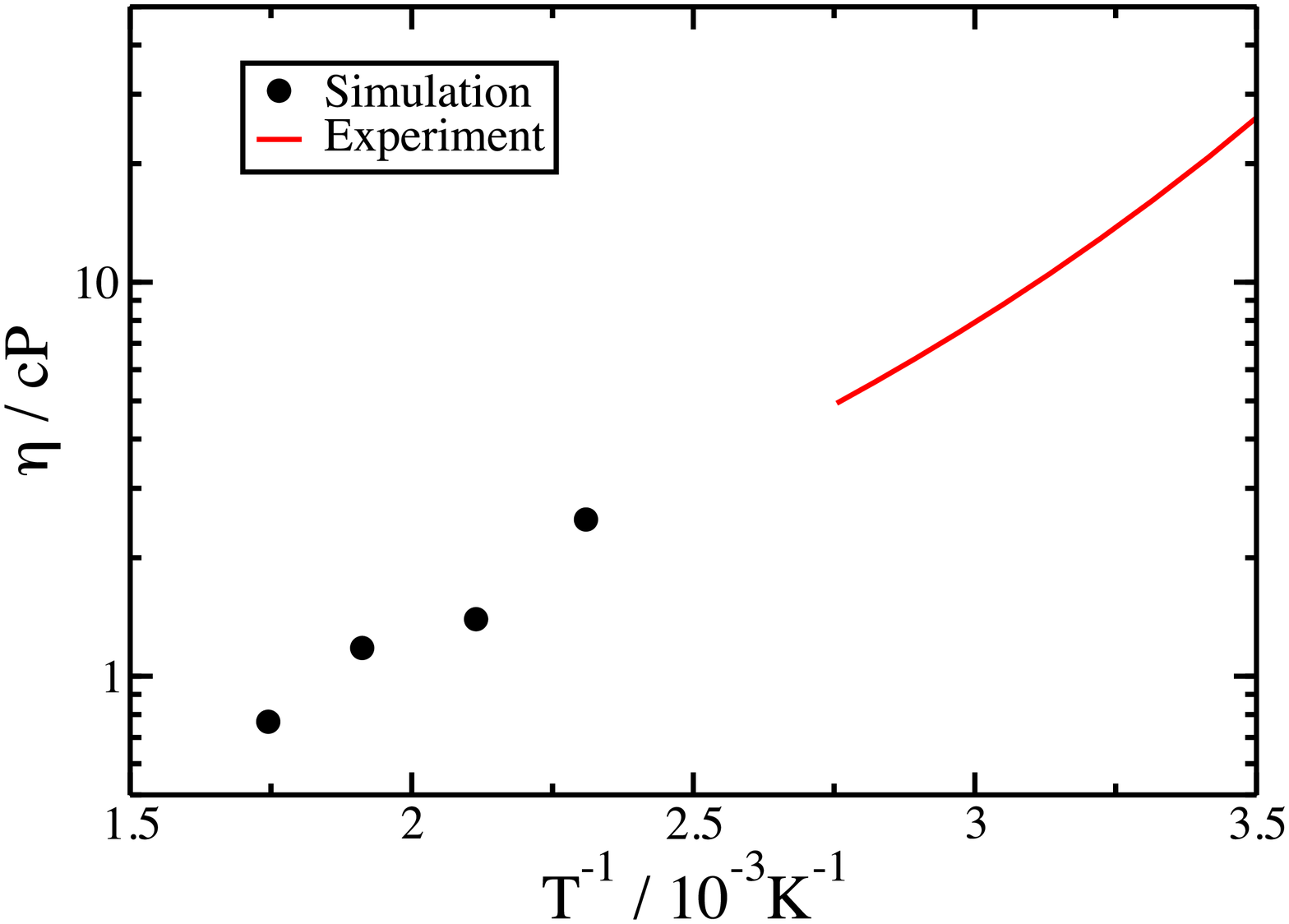}\\\includegraphics[width=.8\columnwidth]{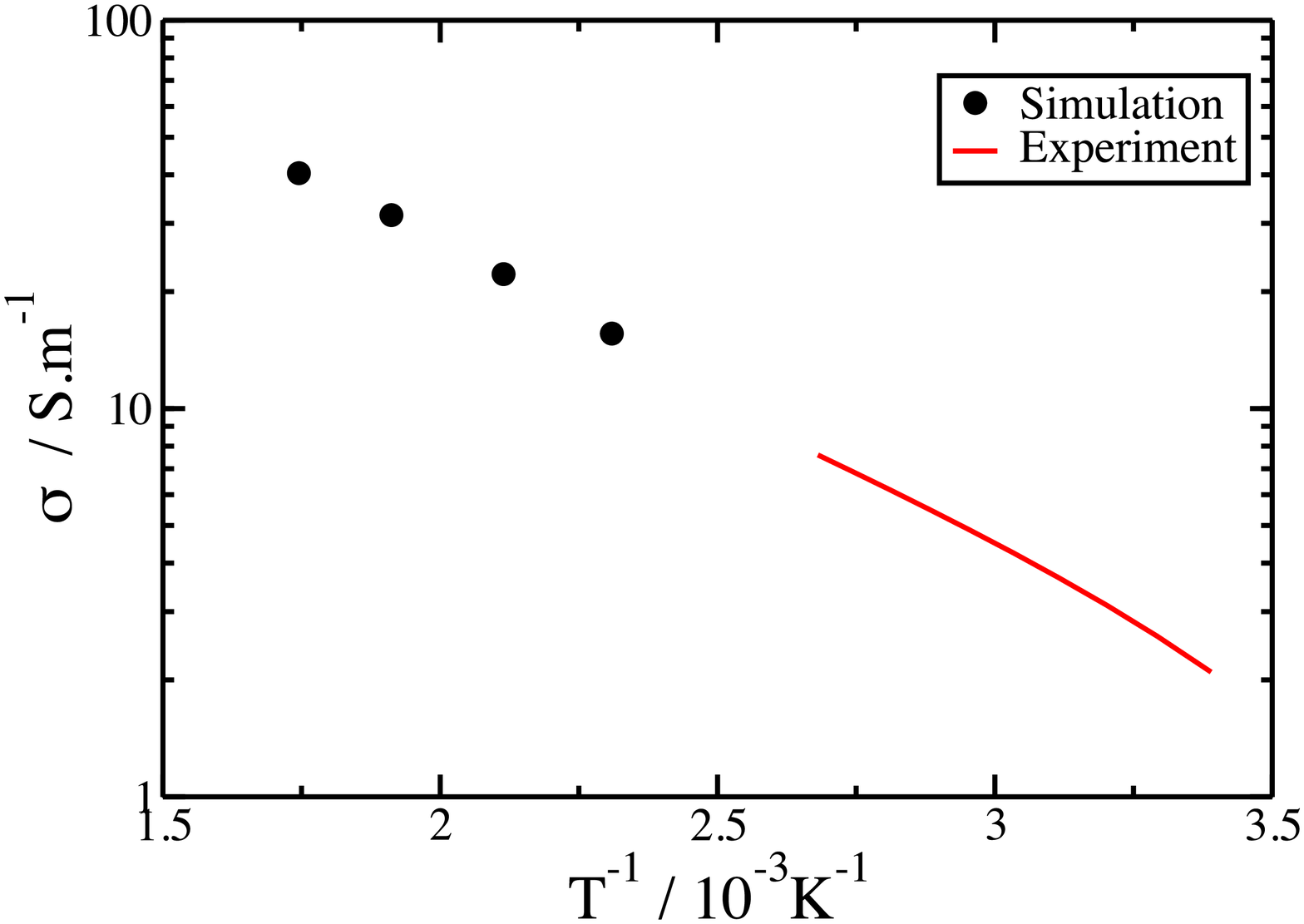}
  \caption{Viscosities (top) and electrical conductivities (bottom) of EMICl-AlCl$_3$ in the form of an Arrhenius plot. The simulated values, although obtained in a different temperature regime, compare well with the experimental ones.~\cite{fannin1984a}}
  \label{fgr:sigmavisco}
\end{center}
\end{figure}

From the mean squared displacement of the overall charge density it is also possible to calculate the electrical conductivity of a melt according to the following formula:

\begin{equation}
\sigma = \frac{e^2}{k_B T V}\lim_{t\rightarrow \infty} \frac{1}{6t}\langle\mid \sum_i q_i \boldsymbol{\delta}{\bf r}_i(t)\mid^2\rangle,
\end{equation}
\noindent where $e$ is the elementary charge and $q_i$ the magnitude of the (partial or formal) charge on atom $i$. Notice that this expression involves the correlations between the displacements of different ions, which is essential in this case as the motion of the Al$^{3+}$ and Cl$^-$ ions within a given AlCl$_4^-$ is strongly correlated. The viscosity is calculated by integration of the correlation function of the stress tensor:
\begin{equation}
\eta=\frac{1}{k_BTV}\int_0^\infty \langle \sigma_{\alpha\beta}(0)\sigma_{\alpha\beta}(\tau)\rangle{\rm d}\tau,
\end{equation}
\noindent where $\sigma_{\alpha\beta}$ is one of the components of the stress tensor. The two latter quantities, which are {\it collective} transport coefficients (compared to the diffusion coefficients, which are {\it individual}), require much longer simulation times to get converged values. Such quantities may therefore only be calculated with the analytic force-field model, they are well out of reach of full ab initio simulations. In the case of EMICl-AlCl$_3$, we could extract their values for all the temperatures except at 298~K. They are shown on figure \ref{fgr:sigmavisco}  in the form of an Arrhenius plot. On the same figure we also show the experimental values;~\cite{fannin1984a} although those were obtained at lower temperatures, we observe a good match between the two sets of data, which confirms that our polarizable ion model yields the correct dynamics of the system.

\begin{table}[h]
  \caption{Diffusion coefficients of the equimolar $M$Cl-AlCl$_3$ systems at 573~K ($M^+$~=~Li$^+$, Na$^+$, K$^+$ or EMI$^+$). All values are in 10$^{-10}$~m$^2$.s$^{-1}$}
  \label{tbl:dynamiccoeff}
  \begin{tabular}{l c c c}
System & $D_{M^+}$ & $D_{\rm Al^{3+}}$ & $D_{\rm Cl^-}$ \\
    \hline
LiCl-AlCl$_3$ & 16.4 & 6.19 & 6.28\\
NaCl-AlCl$_3$ & 20.6 & 6.63 & 6.65\\
KCl-AlCl$_3$ & 24.3 & 8.62& 8.62\\
EMICl-AlCl$_3$ & 35.5 & 23.2 & 23.2\\
    \hline
  \end{tabular}
\end{table}

At the temperature of 573~K, which is above the melting point of the inorganic chloroaluminates, it is also possible to compare the diffusion coefficients of the various ions in all the systems. They are reported in table \ref{tbl:dynamiccoeff}. For all the species, we observe an increase of the diffusion coefficient with the $M^+$ cation size. Note that the number density also changes when switching from one system to another, which is likely to be the main cause for this evolution. Still it is possible to come to the conclusions that in all systems the chloroaluminate anion diffuses more slowly than the corresponding $M^+$ cation. We can also conclude that the organic cation-based ionic liquid, EMICl-AlCl$_3$, fits into the same pattern of behaviour as the inorganic cations-based compounds in regard to the variation of transport properties with cation size.

\section{Conclusion}
We have shown in this study that the polarizable ion model developed for simple molten salts~\cite{ohtori2009a} can successfully be transfered to chloroaluminate ionic liquids. The interaction potential was parameterized purely from first-principles through a force and dipole-matching procedure, and a common set of parameters could be obtained for the mixtures of AlCl$_3$ with several inorganic salts (LiCl, NaCl and KCl) as well as with the organic 1-ethyl-3-methylimidazolium chloride.  An advantage of our model is that it allows us to quantify the chloroaluminate speciation due to the choice of treating Al$^{3+}$ and Cl$^-$ as independent ionic species.

A particular emphasis was given to the study of the equimolar mixture EMICl-AlCl$_3$, for which the structure yielded by the classical simulations performed within the framework of the polarizable ion model is compared to the results obtained from entirely electronic structure-based simulations: An excellent agreement between the two flavors of molecular dynamics is obtained, which enables us to conclude that the structure of EMICl-AlCl$_3$ can almost be viewed as the one of a classical $MX$ molten salt, where $M^+$~=~EMI$^+$ and $X^-$~=~AlCl$_4^-$. The structure is characterized by a competition between close packing and charge ordering effects occuring between the two species. Small deviations from the $MX$ system are observed due to the existence of weak spatial correlations arising from the formation of hydrogen bonds between the ring protons and the chloride anions. When decreasing the cation ionic radius (EMI$^+$~$\rightarrow$~K$^+$~$\rightarrow$~Na$^+$~$\rightarrow$~Li$^+$) small proportions of Al$_2$Cl$_7^-$ due to the reaction
\begin{equation}
{\rm AlCl}_4^- + {\rm AlCl}_4^- \rightarrow {\rm Al}_2{\rm Cl}_7^- + {\rm Cl}^-
\end{equation}
\noindent start to be observed (Al$_3$Cl$_{10}^-$ is also observed in equimolar LiCl-AlCl$_3$). An interesting extension of this work would be to study the same series of systems under non-stoechiometric conditions. The influence of the $M^+$ cation chemical nature in the Lewis acidity of the melt can also be quantified using a recently developed method.~\cite{salanne2011b} In future work we will try to perform such calculations in order to relate it to the chloroaluminate speciation determined in the present study. 
 
The only many-body effect included in the interaction potential for EMICl-AlCl$_3$ is the anion polarization, which enabled us to perform long enough runs to get the dynamic properties of the system, which are out of reach of first-principles simulations. The calculated transport properties were compared to the experimental data, showing again a good agreement for both the individual (diffusion coefficients) and collective (electrical conductivities and viscosities) dynamics of the system.

\section*{Acknowledgments}
MS is grateful to the FAPESP and the Federal University of S\~ao Paulo for a travel grant which enabled this collaboration. LJAS also thanks FAPESP, proc. 08/08670-7.


\end{document}